# Mobility-Dependent and Mobility-Compensated Effective Reproduction Number of COVID-19 Viral Variants: New Metric for Infectivity Evaluation

Sachiko Kodera[1], Essam A. Rashed[1,2], and Akimasa Hirata[1,3]

[1] *Department of Electrical and Mechanical Engineering, Nagoya Institute of Technology, Nagoya 466-8555, Japan*

[2] *Department of Mathematics, Faculty of Science, Suez Canal University, Ismailia 41522, Egypt*

[3] *Center of Biomedical Physics and Information Technology, Nagoya Institute of Technology, Nagoya 466-8555, Japan*

Corresponding Author

Akimasa Hirata, ahirata@nitech.ac.jp    TEL&FAX: +81-52-735-7916

*Department of Electrical and Mechanical Engineering, Nagoya Institute of Technology, Gokiso-cho, Showa-ku, Nagoya, Aichi, 466-8555, Japan*

**Abstract:** During epidemics, estimation of the effective reproduction number (ERN) associated with infectious disease is a challenging topic for policy development and medical resource management. There is still an open question about the dominant factors to characterize in corona virus disease 2019 (COVID-19), although recent studies based on nonlinear regression with machine learning suggested mobility. The emergence of new viral variants is common in widespread pandemics. However, understanding the potential ERN of new variants is required for policy revision, including lockdown constraints. In this study, we proposed time-averaged mobility at transit stations as a surrogate to correlate with ERN using data from three prefectures in Japan. The latency and duration to average over the mobility were 6–8 days and 6–7, respectively ($R^2$ was 0.109–0.512 in Tokyo, 0.365–0.607 in Osaka, and 0.317–0.631 in Aichi). The same linear correlation was confirmed in Singapore and London. The mobility-adjusted ERN of the alpha variant was 15%–30%, and was 20%–40% higher than the standard type in Osaka, Aichi, and London. Similarly, the ERN of the delta variant was 20%–40% higher than that of the standard type in Osaka and Aichi. The proposed metric would be useful for proper evaluation of the infectivity of different variants in terms of ERN.

## 1. Introduction

After the outbreak of the corona virus disease 2019 (COVID-19) (World Health Organization, 2020), everyday routine has been influenced dramatically. In the early period of the COVID-19 pandemic, there were reports of the shortage of medical resources, including its allocation, although some policies, such as city lockdown, have been conducted (Brown et al., 2020, Emanuel et al., 2020). After the discovery and administration of vaccinations, the number of new daily positive cases (DPC) has notably decreased (McKee and Rajan, 2021, Sherman et al., 2021). The first country to reach a high vaccination rate was Israel, which peaked in mid-January 2021. Conversely, its DPC later increased after July and then took another peak in early September (https://www.science.org/content/article/grim-warning-israel-vaccination-blunts-does-not-defeat-delta). This is likely attributable to the wide spread of new delta viral variants (SARS-CoV-2 B.1.617 lineage) whose infectivity would be higher than those of the standard type (Fowlkes et al., 2021, Li et al., 2021).

When new viral variants appeared, their infectivity was one of the concerns in the estimation of medical resource allocation to control potential widespread morbidity. However, its straightforward comparison is difficult as different contributing factors exists, including policy (Carter and May, 2020, Hale et al., 2021), human behavior (Gentili and Cristea, 2020), and the environment (Kodera et al., 2020, Ma et al., 2020, Mecenas et al., 2020, Nazari Harmooshi et al., 2020). A potential useful approach is to transfer knowledge and experience from other countries toward an appropriate and timely policy decision. Several studies investigated the effective reproduction number (ERN) for different factors (Al Wahaibi et al., 2020, Chintalapudi et al., 2020, Xiao et al., 2020), albeit their definitions are not always identical (Arenas et al., 2020, Hellewell et al., 2020, Linka et al., 2020). In addition, due to the incubation time of the infection in addition to the latency between actual infection incident time and reporting in healthcare facilities, estimating the new DPC or ERN is not a straightforward task.

In our previous study, we demonstrated that DPC within future two weeks can be estimated using a machine learning (one of the categories of artificial intelligence) approach with an accuracy of 82.6% (Rashed and Hirata, 2021). The input data were the mobility at different places (retail and recreation, grocery and pharmacy, parks, transit stations, etc.), weather data, and binary labels retributed to other associated factors. However, one drawback of this approach is that the mechanism cannot be explained in a straightforward manner (e.g. nonlinear regression) for easier implementation and data process tracking. The finding in that paper has been supported by a recent study (Kraemer et al., 2020). Thus, a straightforward approach should be investigated following the above findings.

Concerning the situation in Japan, one of the features is that the initial spread of COVID-19 was rather mild and thus vaccination was retarded by a few months as compared to that in European and North American countries. Thus, the effect of vaccination was marginal except those of healthcare professional until June 2021, in which the alpha variant (SARS-CoV-2 B.1.1.7 lineage) was dominant. Therefore, we decided to use the data acquired from different regions in Japan to demonstrate the validity of the proposed metric.

This study aims to further explore a simple surrogate to estimate the ERN from mobility and to discuss the difference of ERN for different viral variants (standard type, and alpha and delta variants). One open question is how much differences are observed between different COVID-19 viral variants for a given public mobility.

## 2. Data and Methods

*2.1. Data*

The three prefectures chosen in this study were Tokyo, Osaka, and Aichi, prefectures whose populations are ranked first, third, and fourth in Japan. The second largest population is Kanagawa, but it is adjacent to Tokyo, and thus we did not consider it. To confirm the tendency in Japan, two other cities were selected: London and Singapore.

Public movements were estimated from Google mobility reports (https://www.google.com/covid19/mobility/ (accessed on 21 May 2021)) that represented global data records from February 15, 2020. On this site, mobility represented the percentage of change from baseline at spots defined as retail and recreation, grocery and pharmacy, parks, transit stations, workplaces, and residential. Baseline is defined as the median value from the 5-week period from January 3, to February 6, 2020.

The numbers of COVID-19 DPC were obtained from the online open data sources at the Japanese Ministry of Health, Labor, and Welfare (https://www.mhlw.go.jp/stf/covid-19/open-data.html (accessed on 14 Oct 2021)).

Figure 1 summarizes new DPC and the mobility at the transit station for three prefectures in Japan. In Japan, there have been five pandemics from February 2020 to October 2021. The stages of the spread were determined to be the same as that in our previous study (Rashed et al., 2020). Table 1 lists the duration of the third (W3), fourth (W4), and fifth waves (W5) in the three prefectures. For each wave, the standard type, alpha variant, and delta variants were dominant, respectively.

*2.2. Effective Reproduction Number*

ERN ($R_t$) data were computed using the following equation:

$$R_t = \left(\sum_{i=1}^{s} DPC_{t-i} \bigg/ \sum_{i=s+1}^{2s} DPC_{t-i}\right)^{\mu/s}, (1)$$

where $s = 7$ is the number of days for a specific time period and $\mu = 5$ (days) is the mean latency after the infection.

*2.3. Averaging Time Windows of Mobility*

In this study, for simplicity, the mobility at the transit stations was considered as a surrogate to represent the transmission of COVID-19 to avoid nonlinear regression. This is based on the finding of our previous study that the most important factor characterizing the new DPC is the mobility at the transit stations (Google mobility). In Japan, the accuracy of the two-week new DPC forecast is >82.6%, whereas the remaining factors would include the weather and the condition of state-of-emergency (Rashed and Hirata, 2021). Thus, we investigated the time windows, which is approximately characterized by the incubation time, to relate with the ERN (Rashed and Hirata, 2021). $R_t$ was averaged from the day corresponding to the latency (days) plus minus a half of the time window (days); if the latency was an even integer, a weighted value was used. The software JMP (SAS Institute, Cary, NC, USA) was used for statistical analysis. In order to specify dominant factors influencing the rates, *p*-value was used. Statistical significance was accepted at $p<0.05$.

## 3. Results

Table 2 shows the relationship between ERN and averaged mobility at the transit station over different durations with different latency setups (e.g., setting the duration to 6 days and latency to 4 days means averaging the mobility 4–9 days before the relevant date). The ERN during long vacations, such as the new year holiday season, summer holidays, consecutive holidays, etc., were excluded because the tendency of the corresponding mobility are different from those of weekdays (Figure 2). As shown in Table 2, the optimal duration and latency were different for different waves. A weaker correlation was observed in Tokyo ($R^2$ was 0.109, 0.512, and 0.235 for W3, W4, and W5, respectively) than those in Osaka ($R^2$ was 0.607, 0.603, and 0.365 for W3, W4, and W5, respectively) and Aichi (0.524, 0.317, and 0.631 for W3, W4, and W5, respectively). For the parameters, the coefficient of correlation was almost the same and statistically significant in the range of 6–8 days for the duration and 6–7 days for the latency. In the following discussion, these values were chosen as 8 and 6 days, respectively, so that averaged $R^2$ becomes maximum for 3 waves in 3 prefectures.

Figure 2 shows the correlation between ERN and average mobility at the transit station wherein the duration and the latency were set to 8 days and 6 days, respectively. Note that the duration affected by the consecutive holidays in July (July 22 to July 25) and Obon holiday (August 7 to August 15), which is a religious holiday in Japan, were excluded when deriving the regression line; the excluded durations were July 27 to August 6 and August 13 to August 28 according to the holidays in July and the Obon holiday, respectively. As shown in Figure 2, a strong correlation between the ERN and the time-averaged mobility at transit stations is confirmed. Although the correlation in Tokyo was worse than that of Osaka and Aichi, a reasonable relation was observed from the Figure. The larger the reduction rate of the mobility at the transit stations, the smaller the ERN.

To confirm the finding in Japanese prefectures, two other international cities were considered:

London and Singapore as shown in Figure 4. Note that the period after the vaccination effect became dominant was excluded. As shown in Figure 4, linear relation was confirmed in both cities.

## 4. Discussion and Summary

Based on our previous finding that mobility is a dominant factor in the determination of DPC with machine learning (nonlinear regression), we explored a surrogate to correlate the ERN in three prefectures of Japan with a linear regression analysis. The motivation for this investigation was that a simple data processing was needed for understanding based on reported new DPC. This nonlinear analysis also prevents straightforward analysis including the effect of viral variants.

Considering the latency, characterized by the incubation time and the delay until the patients reach diagnosis, we have considered the time window for the mobility averaged over, similarly to the ERN. From our analysis, the duration and latency of the mobility were 8 (6–8) and 6 (6–7) days to correlate with the ERN. The latency is in good agreement with the incubation time of 5.1 days (95% confidence interval [CI], 4.5 to 5.8 days) (Lauer et al., 2020) and 5.8 days (95% CI 5.0 to 6.7) (McAloon et al., 2020). The marginal difference may be attributable to the time needed to go to hospital for diagnosis.

The time window is comparable to that of the ERN. The uncertainty of this variation may be attributable to other cofactors including the mobility at different places (e.g., park), environmental conditions, and week day or week end (Rashed and Hirata, 2021). The effect of these other factors on new DPC would be 6.2% on average over six prefectures in Japan. A similar conclusion was reached at in a previous study (Carroll and Prentice, 2021). Note that the correlation for mobility averaged over different durations and latencies and ERN in Tokyo was weaker than that in the remaining prefectures. It was anticipated from our previous study that the estimation accuracy of new DPC in Tokyo would be 62.7% and 83.0% from the mobility at the transit station and those including all places, respectively, even with machine learning (Rashed and Hirata, 2021). On the contrary, these values were comparable to each other, i.e., 75% and 80% in Osaka and Aichi, even only with the mobility at transit stations; significant improvement was observed even including other mobilities. This may be caused by the complexity of mobility in Tokyo where multi core cities or wards exist unlike other prefectures (the population is approximately 12 million).

Using the surrogate identified from Table 2 (latency of 6 days and time window of 8 days), we showed the relationship between the ERN and time-averaged mobility at transit stations for the spread phase of standard type, alpha, and delta variants. As shown in Figure 2, good correlation was observed between these variants especially for Osaka and Aichi. Even at the same reduction rate of the mobility, different ERNs were observed in Osaka and Aichi, suggesting that a target mobility reduction rate is different for policy setting to reduce the mobility to achieve an ERN lower than unity.

Moreover, from Fig. 2, the slope of regression line is close to each other between the standard type, alpha, and delta variants. The difference of the ERN for the alpha variants was 20% higher than that of the standard type in Osaka and Aichi. However, straightforward comparison is not feasible for alpha and delta variants as the rate of fully vaccinated people increases during the spread of delta variants. From Figure 2, the difference of ERN between alpha and delta variants was only a few percent in Osaka whereas 5%–20% in Aichi. Empirically assuming that the effective fully vaccinated population was 20%–25% from July 20 to August 20, compensated ERN in Aichi and Osaka, without considering the breakthrough infection, may be 20%–50% higher than that in the standard type.

This linear tendency was confirmed from London and Singapore for the selected spread duration. From these comparison in addition to the data in Figure 2, the ERN highly depends on the mobility; the mobility reduction to keep the ERN below 1 is different. The ERN for the alpha variants was 20%–40% higher than that of the standard type in London. However, it was not infeasible to calculate the magnitude of the infectivity for the delta variant. According to previous studies in the United Kingdom (Davies et al., 2021, Volz et al., 2021), the infectivity of the delta variant is higher than that of the alpha variant (43%–90%). This tendency is similar to that in Japan, but its magnitude was larger than that in Japan (20%–50% from the standard type; 0%–25% from the alpha variant). A possible reason for this difference would be the mobility adjustment as we proposed in the present study. Spatial variability (Shim et al., 2021) and time variation (Al Wahaibi et al., 2020) have been discussed in earlier studies. These may be attributable to the mobility, as discussed in this study.

In summary, the ERN highly depends on the mobility whose time window is selected. The time window and latency were 8 and 6 days, respectively, from the analysis in three prefectures of Japan. Linear correlation was observed between time-averaged mobility and ERN. For proper comparison of viral infectivity, mobility adjustment is needed. The mobility-adjusted ERN for the alpha and delta variants was 15%–30% and 20%–50% higher than that of the standard type for three prefectures in Japan, which was smaller than but consistent with that observed values in the United Kingdom. This simple metric can provide a useful guideline for a balanced policy enforcement on public movement towards reduction of the viral infectivity with minimum burden on daily-based activities and business.

Table 1: Starting and terminating dates of the spread stages of the third (W3), fourth (W4), and fifth (W5) wave of the COVID-19 pandemic in the three prefectures

|  |  | Spread Duration | | |
|---|---|---|---|---|
| Tokyo | W3 | 8-Nov-20 | - | 4-Jan-21 |
|  | W4 | 11-Mar-21 | - | 3-May-21 |
|  | W5 | 21-Jun-21 | - | 11-Aug-21 |
| Osaka | W3 | 25-Oct-21 | - | 4-Jan-21 |
|  | W4 | 13-Mar-21 | - | 12-Apr-21 |
|  | W5 | 14-Jul-21 | - | 20-Aug-21 |
| Aichi | W3 | 21-Oct-21 | - | 5-Jan-21 |
|  | W4 | 23-Mar-21 | - | 9-May-21 |
|  | W5 | 28-Jul-21 | - | 2-Sep-21 |

Table 2: Coefficient of determination for the correlation between the effective reproduction numbers and the average mobility at transit stations over different durations and latencies in (a) Tokyo, (b) Osaka, and (c) Aichi.

(a)

| Duration | Latency | W3 | | | W4 | | | W5 | | |
|---|---|---|---|---|---|---|---|---|---|---|
| days | days | $R^2$ | p-value | | $R^2$ | p-value | | $R^2$ | p-value | |
| 6 | 4 | 0.098 | 0.017 | (*) | 0.261 | <0.0001 | (***) | 0.005 | 0.666 | (-) |
|  | 5 | 0.109 | 0.011 | (*) | 0.307 | <0.0001 | (***) | 0.059 | 0.121 | (-) |
|  | 6 | 0.079 | 0.032 | (*) | 0.316 | <0.0001 | (***) | 0.119 | 0.025 | (*) |
|  | 7 | 0.035 | 0.158 | (-) | 0.333 | <0.0001 | (***) | 0.160 | 0.009 | (**) |
|  | 8 | 0.006 | 0.552 | (-) | 0.369 | <0.0001 | (***) | 0.208 | 0.002 | (**) |
| 7 | 4 | 0.087 | 0.025 | (*) | 0.286 | <0.0001 | (***) | 0.014 | 0.459 | (-) |
|  | 5 | 0.106 | 0.013 | (*) | 0.334 | <0.0001 | (***) | 0.076 | 0.078 | (-) |
|  | 6 | 0.076 | 0.036 | (*) | 0.343 | <0.0001 | (***) | 0.152 | 0.011 | (*) |
|  | 7 | 0.040 | <0.0001 | (**) | 0.370 | <0.0001 | (***) | 0.190 | 0.004 | (**) |
|  | 8 | 0.010 | <0.0001 | (**) | 0.420 | <0.0001 | (***) | 0.235 | 0.001 | (**) |
| 8 | 4 | 0.085 | 0.026 | (*) | 0.316 | <0.0001 | (***) | 0.024 | 0.325 | (-) |
|  | 5 | 0.102 | 0.015 | (*) | 0.358 | <0.0001 | (***) | 0.103 | 0.039 | (*) |
|  | 6 | 0.077 | 0.035 | (*) | 0.377 | <0.0001 | (***) | 0.177 | 0.006 | (**) |
|  | 7 | 0.043 | 0.120 | (-) | 0.416 | <0.0001 | (***) | 0.213 | 0.002 | (**) |
|  | 8 | 0.021 | 0.280 | (-) | 0.467 | <0.0001 | (***) | 0.204 | 0.003 | (**) |
| 9 | 4 | 0.084 | 0.028 | (*) | 0.341 | <0.0001 | (***) | 0.042 | 0.194 | (-) |

|  | 5 | 0.104 | 0.014 | (*) | 0.387 | <0.0001 | (***) | 0.125 | 0.022 | (*) |
|  | 6 | 0.079 | 0.032 | (*) | 0.418 | <0.0001 | (***) | 0.199 | 0.003 | (**) |
|  | 7 | 0.056 | 0.073 | (-) | 0.458 | <0.0001 | (***) | 0.192 | 0.004 | (**) |
|  | 8 | 0.035 | 0.162 | (-) | 0.512 | <0.0001 | (***) | 0.157 | 0.009 | (**) |

(b)

| Duration | Latency | W3 | | | W4 | | | W5 | | |
| --- | --- | --- | --- | --- | --- | --- | --- | --- | --- | --- |
| days | days | $R^2$ | p-value | | $R^2$ | p-value | | $R^2$ | p-value | |
| 6 | 4 | 0.518 | <0.0001 | (***) | 0.523 | <0.0001 | (***) | 0.344 | 0.007 | (**) |
|  | 5 | 0.587 | <0.0001 | (***) | 0.446 | <0.0001 | (***) | 0.298 | 0.013 | (*) |
|  | 6 | 0.584 | <0.0001 | (***) | 0.357 | 0.0004 | (***) | 0.312 | 0.011 | (*) |
|  | 7 | 0.517 | <0.0001 | (***) | 0.276 | 0.002 | (**) | 0.232 | 0.032 | (*) |
|  | 8 | 0.438 | <0.0001 | (***) | 0.227 | 0.007 | (**) | 0.110 | 0.154 | (-) |
| 7 | 4 | 0.551 | <0.0001 | (***) | 0.603 | 0.000 | (***) | 0.365 | 0.005 | (**) |
|  | 5 | 0.607 | <0.0001 | (***) | 0.478 | <0.0001 | (***) | 0.346 | 0.006 | (**) |
|  | 6 | 0.596 | <0.0001 | (***) | 0.379 | 0.0002 | (***) | 0.292 | 0.014 | (*) |
|  | 7 | 0.533 | <0.0001 | (***) | 0.316 | 0.001 | (**) | 0.243 | 0.027 | (*) |
|  | 8 | 0.463 | <0.0001 | (***) | 0.291 | 0.002 | (**) | 0.200 | 0.048 | (*) |
| 8 | 4 | 0.566 | <0.0001 | (***) | 0.510 | <0.0001 | (***) | 0.362 | 0.005 | (**) |
|  | 5 | 0.605 | <0.0001 | (***) | 0.415 | <0.0001 | (***) | 0.265 | 0.020 | (*) |
|  | 6 | 0.587 | <0.0001 | (***) | 0.353 | 0.0004 | (***) | 0.216 | 0.039 | (*) |
|  | 7 | 0.532 | <0.0001 | (***) | 0.324 | 0.001 | (**) | 0.229 | 0.033 | (*) |
|  | 8 | 0.477 | <0.0001 | (***) | 0.302 | 0.001 | (**) | 0.172 | 0.069 | (-) |
| 9 | 4 | 0.575 | <0.0001 | (***) | 0.437 | <0.0001 | (***) | 0.309 | 0.011 | (*) |
|  | 5 | 0.599 | <0.0001 | (***) | 0.380 | 0.0002 | (***) | 0.212 | 0.041 | (*) |
|  | 6 | 0.581 | <0.0001 | (***) | 0.350 | 0.001 | (**) | 0.191 | 0.054 | (-) |
|  | 7 | 0.537 | <0.0001 | (***) | 0.327 | 0.001 | (***) | 0.176 | 0.0004 | (***) |
|  | 8 | 0.486 | <0.0001 | (***) | 0.291 | 0.002 | (**) | 0.151 | 0.090 | (-) |

(c)

|  |  | W3 | | | W4 | | | W5 | | |
| --- | --- | --- | --- | --- | --- | --- | --- | --- | --- | --- |
| Duration | Latency | $R^2$ | p-value | | $R^2$ | p-value | | $R^2$ | p-value | |
| 6 | 4 | 0.330 | <0.0001 | (***) | 0.263 | 0.0002 | (***) | 0.416 | 0.032 | (*) |
|  | 5 | 0.393 | <0.0001 | (***) | 0.213 | 0.001 | (***) | 0.322 | 0.069 | (-) |
|  | 6 | 0.454 | <0.0001 | (***) | 0.214 | 0.001 | (***) | 0.337 | 0.061 | (-) |
|  | 7 | 0.468 | <0.0001 | (***) | 0.256 | 0.000 | (***) | 0.426 | 0.030 | (*) |

|   |   |   |   |   |   |   |   |   |   |
|---|---|---|---|---|---|---|---|---|---|
|   | 8 | 0.442 | <0.0001 | (***) | 0.278 | <0.0001 | (***) | 0.489 | 0.017 | (*) |
|   | 4 | 0.350 | <0.0001 | (***) | 0.260 | 0.0002 | (***) | 0.475 | 0.019 | (*) |
|   | 5 | 0.426 | <0.0001 | (***) | 0.225 | 0.001 | (**) | 0.481 | 0.018 | (*) |
| 7 | 6 | 0.484 | <0.0001 | (***) | 0.277 | 0.0001 | (***) | 0.472 | 0.020 | (*) |
|   | 7 | 0.492 | <0.0001 | (***) | 0.317 | <0.0001 | (***) | 0.440 | 0.026 | (*) |
|   | 8 | 0.485 | <0.0001 | (***) | 0.307 | <0.0001 | (***) | 0.448 | 0.024 | (*) |
|   | 4 | 0.373 | <0.0001 | (***) | 0.268 | <0.0001 | (***) | 0.577 | 0.007 | (**) |
|   | 5 | 0.446 | <0.0001 | (***) | 0.263 | 0.0002 | (***) | 0.566 | 0.008 | (**) |
| 8 | 6 | 0.492 | <0.0001 | (***) | 0.302 | <0.0001 | (***) | 0.473 | 0.019 | (*) |
|   | 7 | 0.508 | <0.0001 | (***) | 0.303 | <0.0001 | (***) | 0.389 | 0.040 | (*) |
|   | 8 | 0.504 | <0.0001 | (***) | 0.261 | 0.0002 | (***) | 0.273 | 0.099 | (-) |
|   | 4 | 0.394 | <0.0001 | (***) | 0.299 | <0.0001 | (***) | 0.631 | 0.004 | (**) |
|   | 5 | 0.460 | <0.0001 | (***) | 0.282 | <0.0001 | (***) | 0.556 | 0.008 | (**) |
| 9 | 6 | 0.510 | <0.0001 | (***) | 0.294 | <0.0001 | (***) | 0.417 | 0.032 | (*) |
|   | 7 | 0.524 | <0.0001 | (***) | 0.266 | 0.0001 | (***) | 0.229 | 0.137 | (-) |
|   | 8 | 0.519 | <0.0001 | (***) | 0.216 | 0.001 | (***) | 0.130 | 0.277 | (-) |

*: $p < 0.05$, **: $p < 0.01$, ***: $p < 0.001$

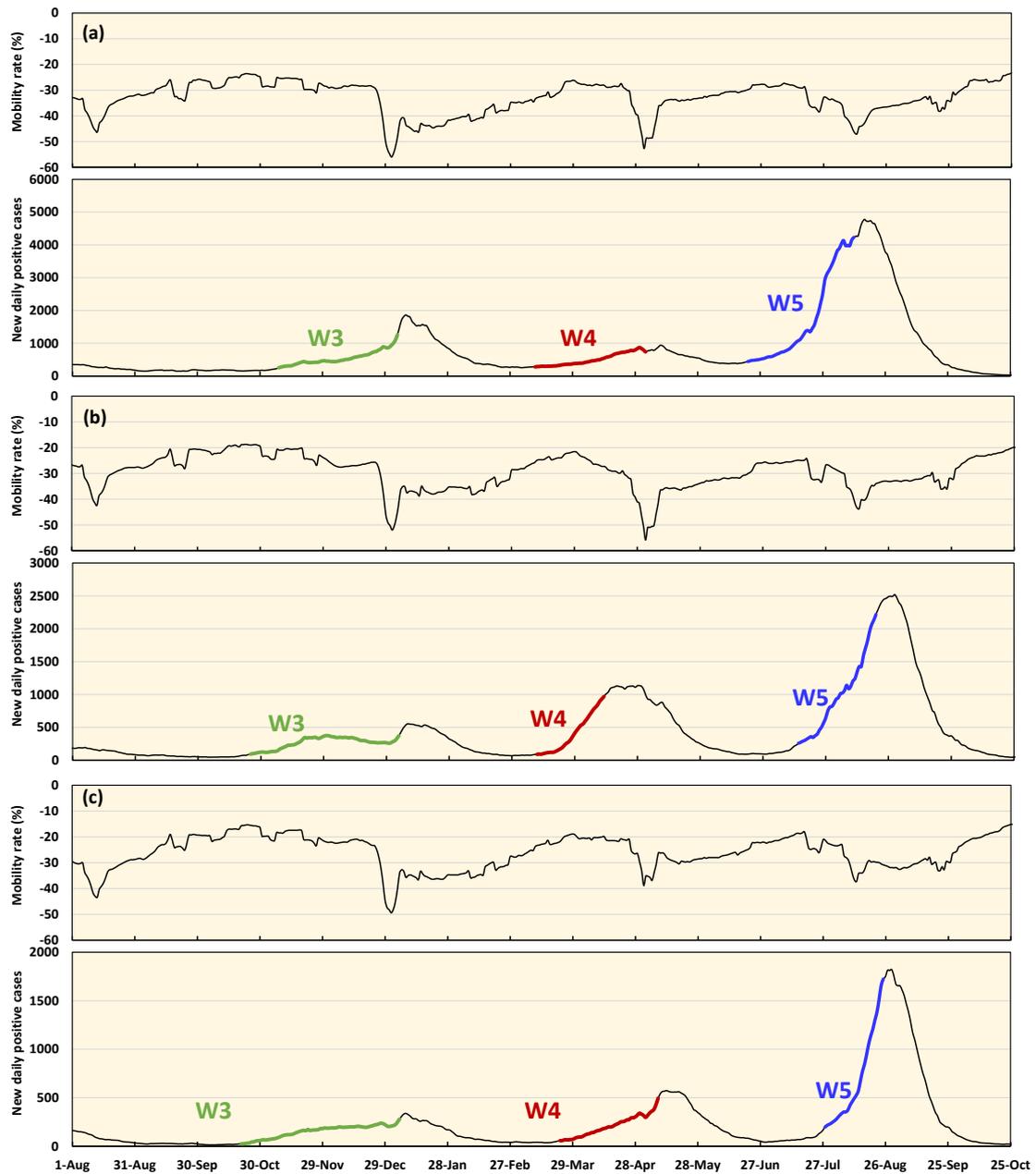

**Figure 1:** (a–c) Daily confirmed new positive cases and (d–f) mobility change (%) at transit stations in (a, d) Tokyo, (b, e) Osaka, and (c, f) Aichi prefectures. The colored lines show the spread duration of the third, fourth, and fifth pandemic waves. All lines represent seven-day averages.

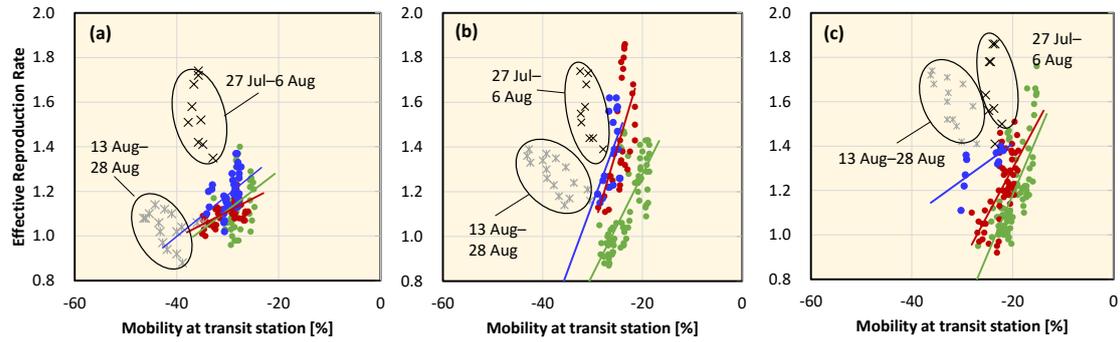

**Figure 2:** Correlation between mobility at the transit station and effective reproduction rate in (a) Tokyo, (b) Osaka, and (c) Aichi. Mobility was averaged over 8 days with a latency of 6 days. Green, red, and blue colors correspond to the third, fourth, and fifth waves shown in Figure 1, respectively.

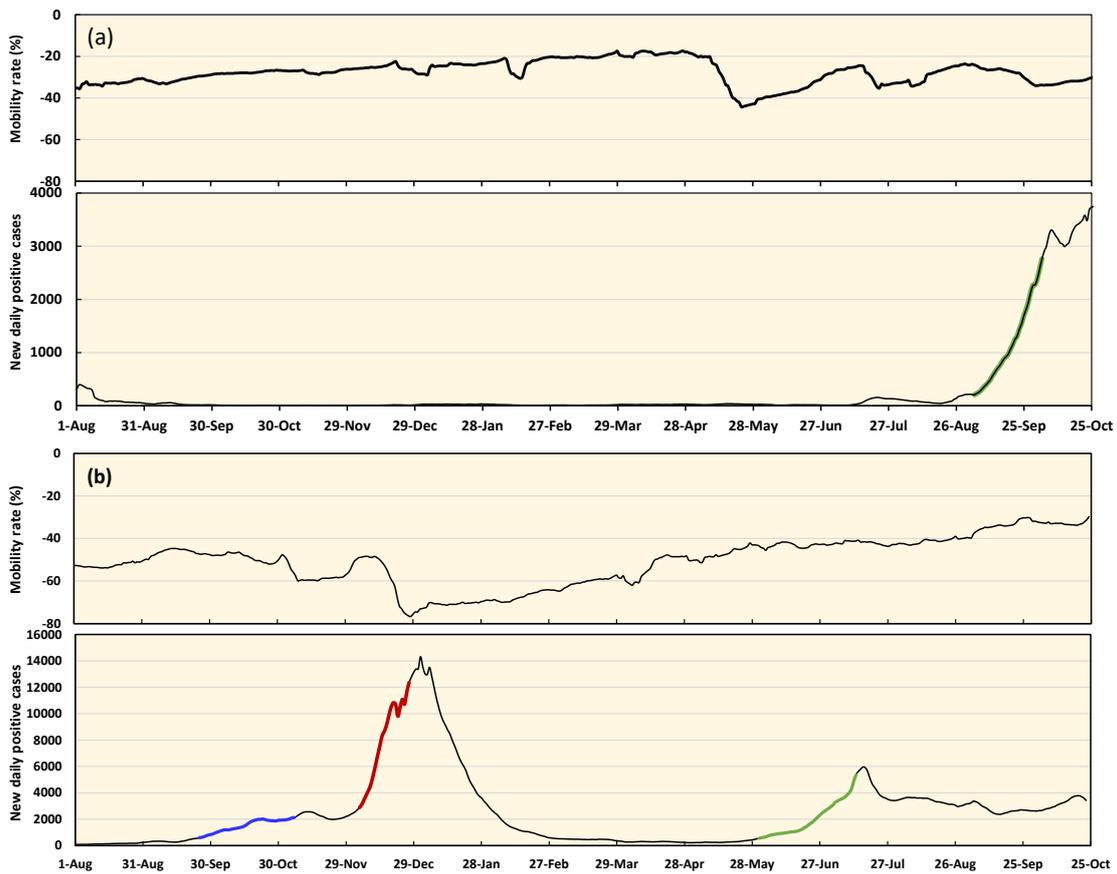

**Figure 3:** Mobility and new daily confirmed cases in (a) Singapore and (b) London.

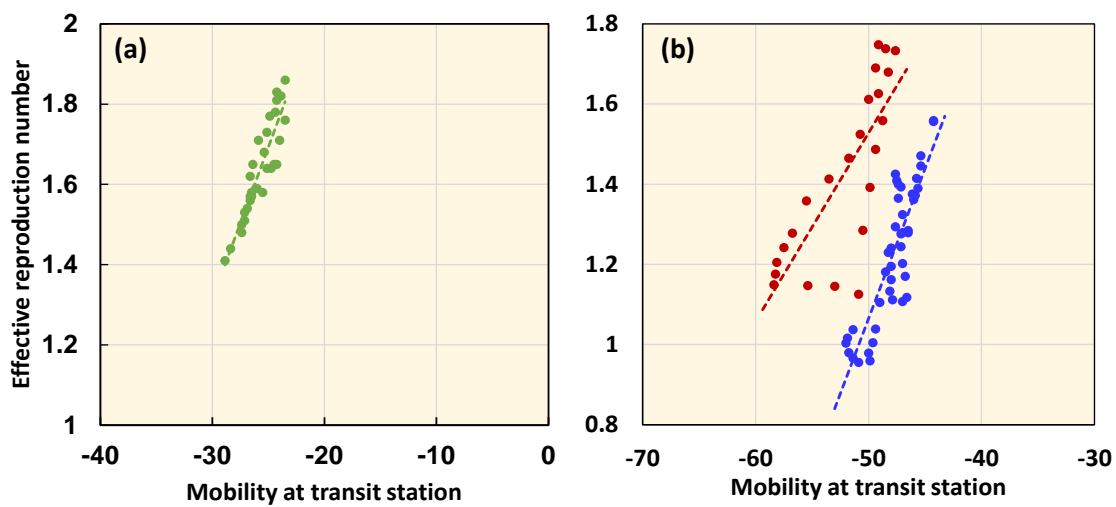

**Figure 4:** Correlation between mobility at transit stations and effective reproduction rate in (a) Singapore and (b) London. Mobility was averaged over 8 days with a latency of 6 days.

## References

Al Wahaibi A, Al Manji A, Al Maani A, Al Rawahi B, Al Harthy K, Alyaquobi F, et al. COVID-19 epidemic monitoring after non-pharmaceutical interventions: The use of time-varying reproduction number in a country with a large migrant population. International Journal of Infectious Diseases 2020;99:466-72.

Arenas A, Cota W, Gomez-Gardenes J, Gómez S, Granell C, Matamalas JT, et al. Derivation of the effective reproduction number R for COVID-19 in relation to mobility restrictions and confinement. medRxiv 2020.

Brown MJ, Goodwin J, Liddell K, Martin S, Palmer S, Firth P, et al. Allocating medical resources in the time of Covid-19. N Engl J Med 2020;382:e79.

Carroll R, Prentice CR. Community vulnerability and mobility: What matters most in spatio-temporal modeling of the COVID-19 pandemic? Social Science & Medicine 2021;287:114395.

Carter DP, May PJ. Making sense of the US COVID-19 pandemic response: A policy regime perspective. Administrative Theory & Praxis 2020;42(2):265-77.

Chintalapudi N, Battineni G, Sagaro GG, Amenta F. COVID-19 outbreak reproduction number estimations and forecasting in Marche, Italy. International Journal of Infectious Diseases 2020;96:327-33.

Davies NG, Abbott S, Barnard RC, Jarvis CI, Kucharski AJ, Munday JD, et al. Estimated transmissibility and impact of SARS-CoV-2 lineage B. 1.1. 7 in England. Science 2021;372(6538).

Emanuel EJ, Persad G, Upshur R, Thome B, Parker M, Glickman A, et al. Fair allocation of scarce medical resources in the time of Covid-19. Mass Medical Soc; 2020.

Fowlkes A, Gaglani M, Groover K, Thiese MS, Tyner H, Ellingson K, et al. Effectiveness of COVID-19 vaccines in preventing SARS-CoV-2 infection among frontline workers before and during B. 1.617. 2

(Delta) variant predominance—eight US locations, December 2020–August 2021. Morbidity and Mortality Weekly Report 2021;70(34):1167.

Gentili C, Cristea IA. Challenges and opportunities for human behavior research in the coronavirus disease (COVID-19) pandemic. Frontiers in psychology 2020;11:1786.

Hale T, Angrist N, Goldszmidt R, Kira B, Petherick A, Phillips T, et al. A global panel database of pandemic policies (Oxford COVID-19 Government Response Tracker). Nature Human Behaviour 2021;5(4):529-38.

Hellewell J, Abbott S, Gimma A, Bosse NI, Jarvis CI, Russell TW, et al. Feasibility of controlling COVID-19 outbreaks by isolation of cases and contacts. The Lancet Global Health 2020;8(4):e488-e96.

Kodera S, Rashed EA, Hirata A. Correlation Between COVID-19 Morbidity and Mortality Rates in Japan and Local Population Density, Temperature, and Absolute Humidity. International Journal of Environmental Research and Public Health 2020;17(15):5477.

Kraemer MU, Yang C-H, Gutierrez B, Wu C-H, Klein B, Pigott DM, et al. The effect of human mobility and control measures on the COVID-19 epidemic in China. Science 2020;368(6490):493-7.

Lauer SA, Grantz KH, Bi Q, Jones FK, Zheng Q, Meredith HR, et al. The incubation period of coronavirus disease 2019 (COVID-19) from publicly reported confirmed cases: estimation and application. Annals of internal medicine 2020;172(9):577-82.

Li B, Deng A, Li K, Hu Y, Li Z, Xiong Q, et al. Viral infection and transmission in a large well-traced outbreak caused by the Delta SARS-CoV-2 variant. MedRxiv 2021.

Linka K, Peirlinck M, Kuhl E. The reproduction number of COVID-19 and its correlation with public health interventions. Computational Mechanics 2020;66(4):1035-50.

Ma Y, Zhao Y, Liu J, He X, Wang B, Fu S, et al. Effects of temperature variation and humidity on the death of COVID-19 in Wuhan, China. Science of The Total Environment 2020:138226.

McAloon C, Collins Á, Hunt K, Barber A, Byrne AW, Butler F, et al. Incubation period of COVID-19: a rapid systematic review and meta-analysis of observational research. BMJ open 2020;10(8):e039652.

McKee M, Rajan S. What can we learn from Israel's rapid roll out of COVID 19 vaccination? Israel journal of health policy research 2021;10(1):1-4.

Mecenas P, Bastos RTdRM, Vallinoto ACR, Normando D. Effects of temperature and humidity on the spread of COVID-19: A systematic review. PLoS one 2020;15(9):e0238339.

Nazari Harmooshi N, Shirbandi K, Rahim F. Environmental concern regarding the effect of humidity and temperature on SARS-COV-2 (COVID-19) survival: fact or fiction. Kiarash and Rahim, Fakher, Environmental Concern Regarding the Effect of Humidity and Temperature on SARS-COV-2 (COVID-19) Survival: Fact or Fiction (March 29, 2020) 2020.

Rashed EA, Hirata A. Infectivity Upsurge by COVID-19 viral variants in Japan: Evidence from deep learning modeling. International Journal of Environmental Research and Public Health 2021;18(15):7799.

Rashed EA, Kodera S, Gomez-Tames J, Hirata A. Influence of absolute humidity, temperature and population density on COVID-19 spread and decay durations: multi-prefecture study in Japan. International Journal of Environmental Research and Public Health 2020;17(15):5354.

Sherman SM, Smith LE, Sim J, Amlôt R, Cutts M, Dasch H, et al. COVID-19 vaccination intention in the UK: results from the COVID-19 vaccination acceptability study (CoVAccS), a nationally representative cross-sectional survey. Human vaccines & immunotherapeutics 2021;17(6):1612-21.

Shim E, Tariq A, Chowell G. Spatial variability in reproduction number and doubling time across two waves of the COVID-19 pandemic in South Korea, February to July, 2020. International Journal of Infectious Diseases 2021;102:1-9.

Volz E, Mishra S, Chand M, Barrett JC, Johnson R, Geidelberg L, et al. Transmission of SARS-CoV-2 Lineage B. 1.1. 7 in England: Insights from linking epidemiological and genetic data. MedRxiv 2021:2020.12. 30.20249034.

World Health Organization. Coronavirus disease 2019 (COVID-19): situation report, 72. 2020.

Xiao Y, Tang B, Wu J, Cheke RA, Tang S. Linking key intervention timing to rapid decline of the COVID-19 effective reproductive number to quantify lessons from mainland China. International Journal of Infectious Diseases 2020;97:296-8.